\documentclass[a4paper,11pt]{article}
\usepackage{pos}
\usepackage{amsmath}
\usepackage{float}
\usepackage{gensymb}
\usepackage{caption}
\usepackage{subcaption}

\title{POCAM in the IceCube Upgrade}
 \ShortTitle{}

\author{The IceCube Collaboration \\{\normalsize \normalfont(a complete list of authors can be found at the end of the proceedings)}}
\emailAdd{nikhita.khera@tum.de}
\emailAdd{felix.henningsen@tum.de}

\abstract{The IceCube Neutrino Observatory at the geographic South Pole instruments a gigaton of glacial Antarctic ice with over 5000 photosensors. The detector, by now running for over a decade, will be upgraded with seven new densely instrumented strings. The project focuses on the improvement of low-energy and oscillation physics sensitivities as well as re-calibration of the existing detector. Over the last few years we developed a Precision Optical Calibration Module (POCAM) providing self-monitored, isotropic, nanosecond, light pulses for optical calibration of large-volume detectors. Over 20 next-generation POCAMs will be calibrated and deployed in the IceCube Upgrade in order to reduce existing detector systematics. We report a general overview of the POCAM instrument, its performance and calibration procedures.\\

\vspace{4mm}
{\bfseries Corresponding authors:}
Nikhita Khera$^{1 *}$, Felix Henningsen$^{1,2}$\\
{$^{1}$ \itshape Technical University of Munich, James-Frank-Str. 1, 85748 Garching, Germany}\\
{$^{2}$ \itshape Max Planck Institute for Physics, F\"ohringer Ring 6, 80805 M\"unchen, Germany}\\[4mm]
$^*$ Presenter

\FullConference{37$^{\rm{th}}$ International Cosmic Ray Conference (ICRC 2021)\\
		July 12th -- 23rd, 2021\\
		Online -- Berlin, Germany}

}

\begin{document}
\maketitle
\section{POCAM and its housing}
\begin{figure}[H]
	\centering
	\includegraphics[height=5cm, width=8cm]{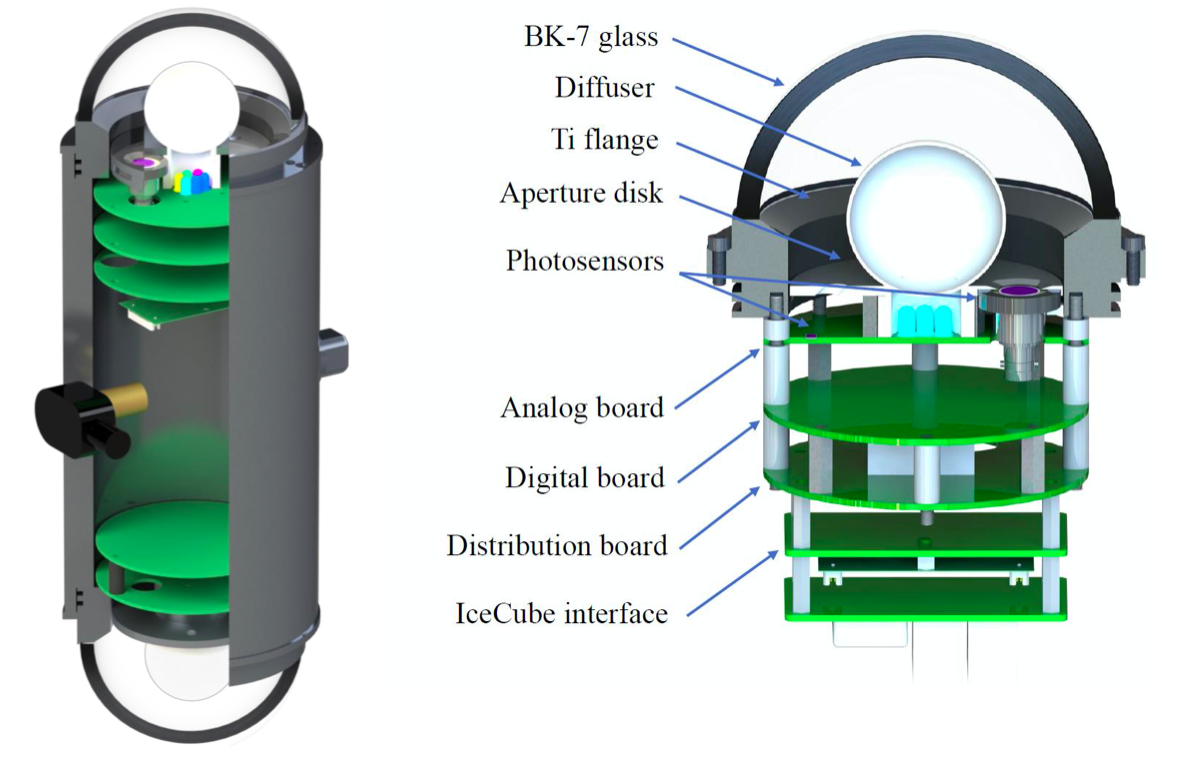}
	\caption{\textit{POCAM module and POCAM hemisphere assembly \cite{3abc}}}
	\label{fig1}
	\end{figure}

The Precision Optical Calibration Module (POCAM) \cite{1abc} \cite{2abc} is designed to be a high precision, self-monitoring, isotropic, nanosecond, multi-wavelength calibration light source for a precise measurement of the energy scale and resolution within the IceCube Upgrade \cite{9abc}. 
This instrument, as shown in Figure \ref{fig1}, has a cylindrical titanium housing, which hosts the circuit boards with light flashers, read out electronics and a PTFE integrating sphere, encapsulated finally by a BK-7 glass hemisphere. Two such identical hemisphere assemblies are on either side of the titanium cylinder. The analog PCB has six optical emitters - three light emitting diodes (LEDs) (365nm, 405nm, 465 nm) and three laser diodes (LDs) (405nm, 455nm, 520 nm)- for flashing. On top of these, integrated on the aperture disk, are two sensors - a silicon photomultiplier (SiPM) and a photodiode (PD) - installed for self-monitoring. The PTFE integrating sphere is responsible for making the light pulse isotropic \cite{3abc} \cite{8abc}.

To verify the performance and stability of the POCAM,  the housing was tested in a high pressure chamber at Nautilus and temperature testing was done by placing it in a freezer at  -55 \degree C for five weeks. Vibration and shock tests on the POCAM stack (i.e. the hemisphere assembly as shown in Figure \ref{fig1} without the glass) were performed and passed at the IABG test site in Ottobrunn, Germany.
After the development of the POCAM started in the late 2014, a first protype of the device was successfully deployed in the GVD neutrino telescope at Lake Baikal, Russia in 2017 \cite{1abc}.  An improved second iteration of the device was then installed within the STRAW experiment \cite{4abc} in the North East Pacific deep sea in the year 2018. After these succesfull deployments, this latest iteration of the POCAM that we report about in this paper, with significant improvements and changes, is now being developed in the scope of the IceCube Upgrade project \cite{3abc} \cite{5abc}.

\section{Light Emitters}
\begin{figure}[H]
	\centering
	\includegraphics[height=3.5cm, width=14cm]{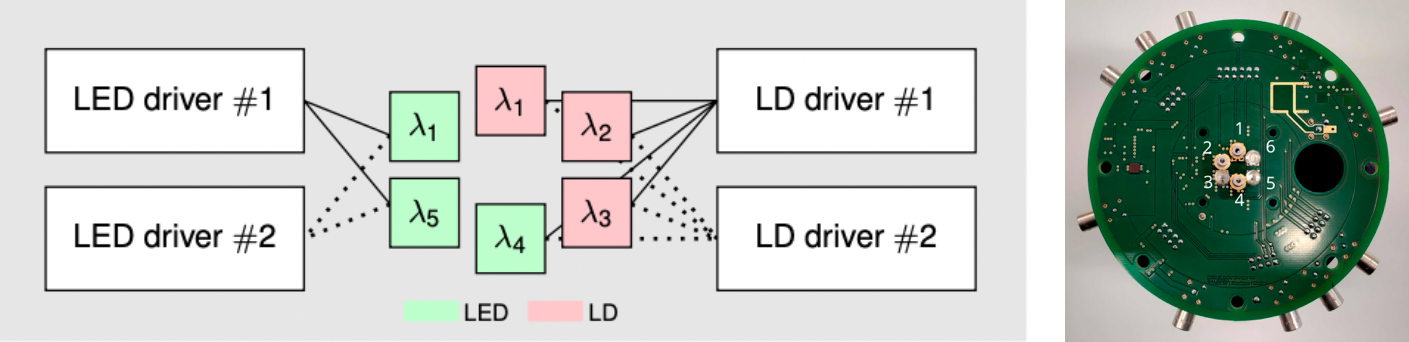}
	\caption{\textit{Layout of LEDs, LDs and pulse drivers in the POCAM (left) \cite{3abc}, LEDs and LDs on the analog board (right)}}
	\label{fig2}
	\end{figure}

The analog boards in the POCAM electronics incorporate six light emitters out of which three are LEDs and three are LDs. Further, two Kapustinsky \cite{6abc} and two LD-type LD drivers \cite{7abc} are responsible for the driving of light pulses from these emitters (Figure \ref{fig2}). The purpose to keep two of each type of drivers is to not only allow redundancy but also a freedom to selectively drive different pulsers for different flasher wavelengths. We have two LD type drivers as well as a fast and a default Kapustinsky circuit. 

\begin{figure}[H]
	\centering
	\includegraphics[height=5.3cm, width=14.4cm]{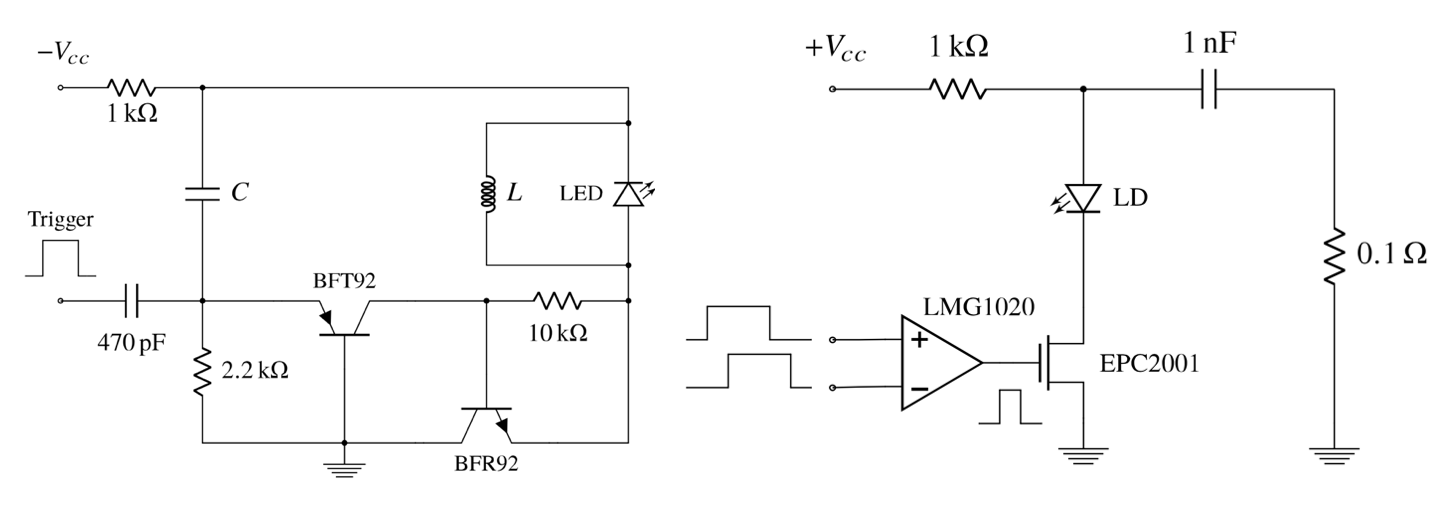}
	\caption{\textit{Schematic of the Kapustinsky circuit (left) and the LD type driver circuit (right) \cite{3abc}}}
	\label{fig3}
	\end{figure}

The Kapustinsky circuit is developed for LEDs and makes use of a relatively simple and pre-proven circuitry as shown in the Figure \ref{fig3}. The circuit operates on negative bias voltage and produces a pulse when triggered with a square pulse signal. The pulse properties can be modified by controlling the values of capacitor C and the inductance L and the pulse width scales qualitatively with $\sqrt{LC}$. 

For the POCAM application in IceCube Upgrade, the Kapustinsky circuit is configured to drive the 405 nm and 465 nm LEDs. Each of these LEDs can select a fast (L, C - 22 nH, 100 pF) or default (L,C - 22 nH, 1.2 nF) Kapustinsky driver configuration with different pulse widths and light yields.
	
The third and the fourth driver on the analog board are the LD type drivers which are configured to drive a 365 nm LED and three LDs at 405 nm, 455 nm and 520 nm. These drivers are equipped with fast high-current switching GaN-FETS (Gallium-Nitride field effect transistors). The picosecond switching of these GaN-FETs allow the possibility of adjustable pulse widths and significantly more light output. The circuitry of the LD driver is also shown in the Figure \ref{fig3}. The circuit operates on positive bias voltage and produces a pulse proportional to the input signal.

\begin{figure}[H]
	\centering
	\includegraphics[height=5cm, width=11cm]{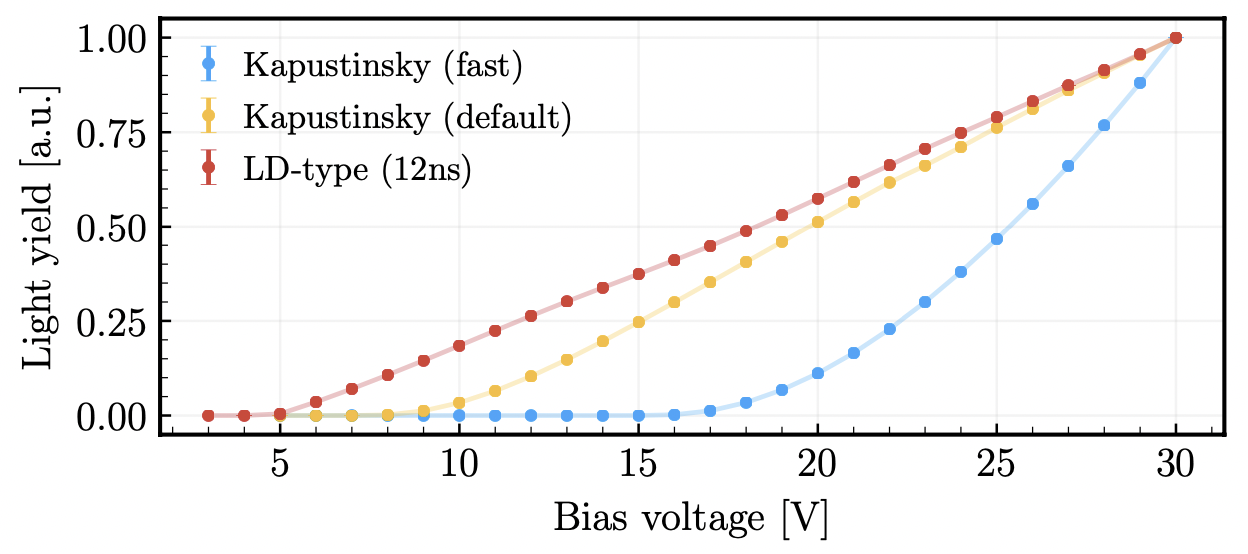}
	\caption{\textit{Intensity yield (measured with an external photodiode) as a function of voltage for all the drivers at 405 nm. \cite{3abc}}}
	\label{fig4}
	\end{figure}	
	
\begin{figure}[H]
	\centering
	\includegraphics[height=5cm, width=11cm]{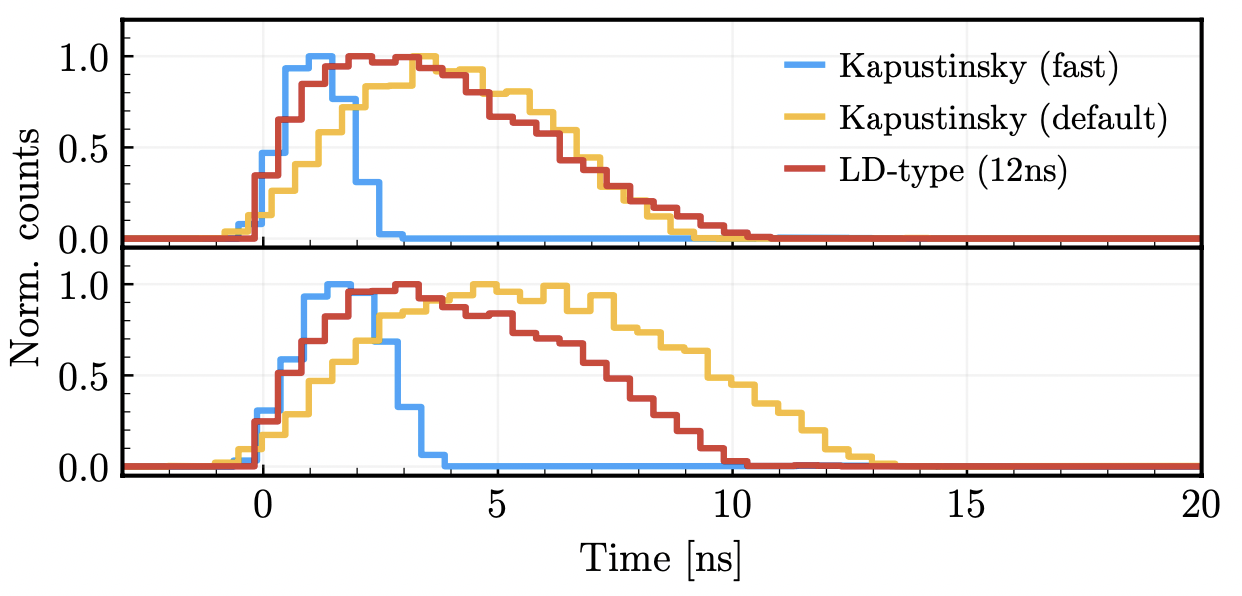}
	\caption{\textit{Time profiles (measured using an avalanche photodidoe) at minimal-working (top) and maximal (bottom) applied bias voltage for all the drivers at 405 nm.\cite{3abc}}}
	\label{fig5}
	\end{figure}

Both flasher circuits perform in different regions of the dynamic range of the POCAM. The intensity behaviour of all driver circuits (observed using an external photodiode) and their respective time profiles (using an avalanche photodiode) were measured for the 405nm LED.  These can be seen in Figures \ref{fig4} and \ref{fig5} respectively. 

As visible in the voltage vs. light yield plot (Figure \ref{fig4}), the fastest Kapustinsky has a significantly late light onset, followed by the default Kapustinsky and then the LD driver which shows linear intensity behaviour. The timing profiles (Figure \ref{fig5}) for the Kapustinskies have been fine tuned to an FWHM of 1 - 3 ns for the fast and 4 - 8 ns for the default configuration. The LD drivers on the other hand are optimized to generate a wide range of pulse widths (here 12ns).

\section{Self-monitoring sensors}
In order for the POCAM to self monitor the light output per pulse and correct for any intensity fluctuations, two sensors- a SiPM and a photodiode (PD) - are embedded into the aperture disk of the POCAM. The SiPM intends to work in the low intensity regime  and makes use of an FPGA discriminator, the signal from which is then fed into a time-digital-converter (TDC) to determine the time-over-threshold (ToT). Whereas the photodiode is responsible for high intensity light measurements and makes use of a transimpedence amplifier (TIA), followed by secondary high gain and low gain amplifiers, which provides a voltage amplitude proportional to the measured charge of the PD.

\begin{figure}[H]
\begin{subfigure}{.49\textwidth}
  \centering
  % include first image
  \includegraphics[height=3.5cm, width=7cm]{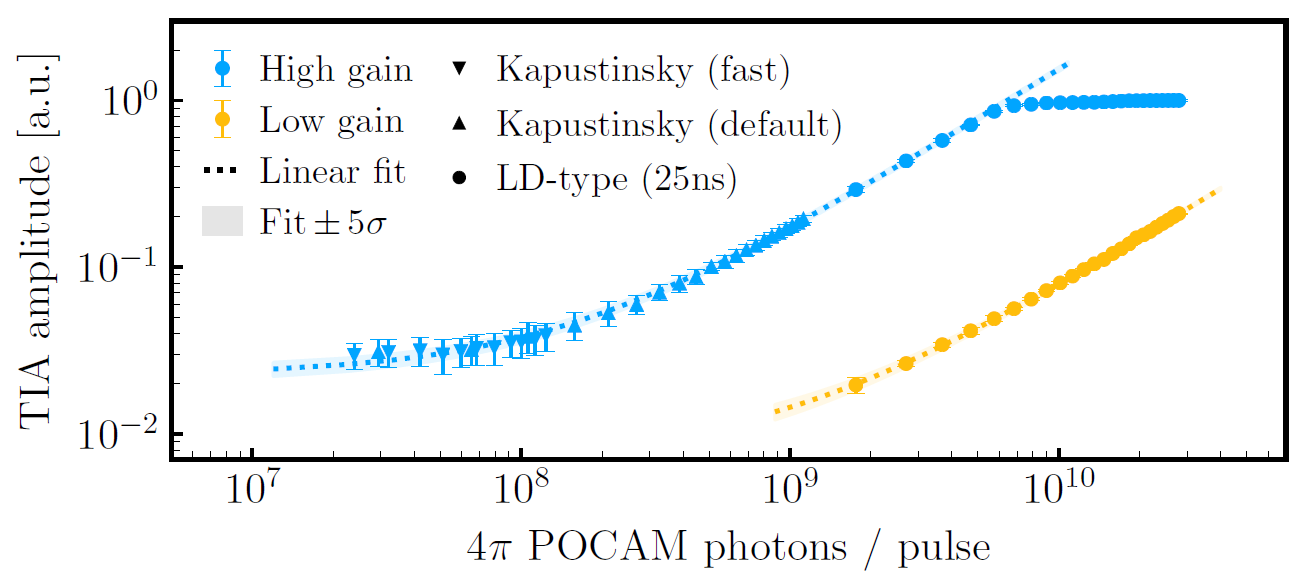}\hspace{1em}
%  \label{fig:sub-first}
\end{subfigure}
\begin{subfigure}{.37\textwidth}
  \centering
  % include second image
  \includegraphics[height=3.5cm, width=7cm]{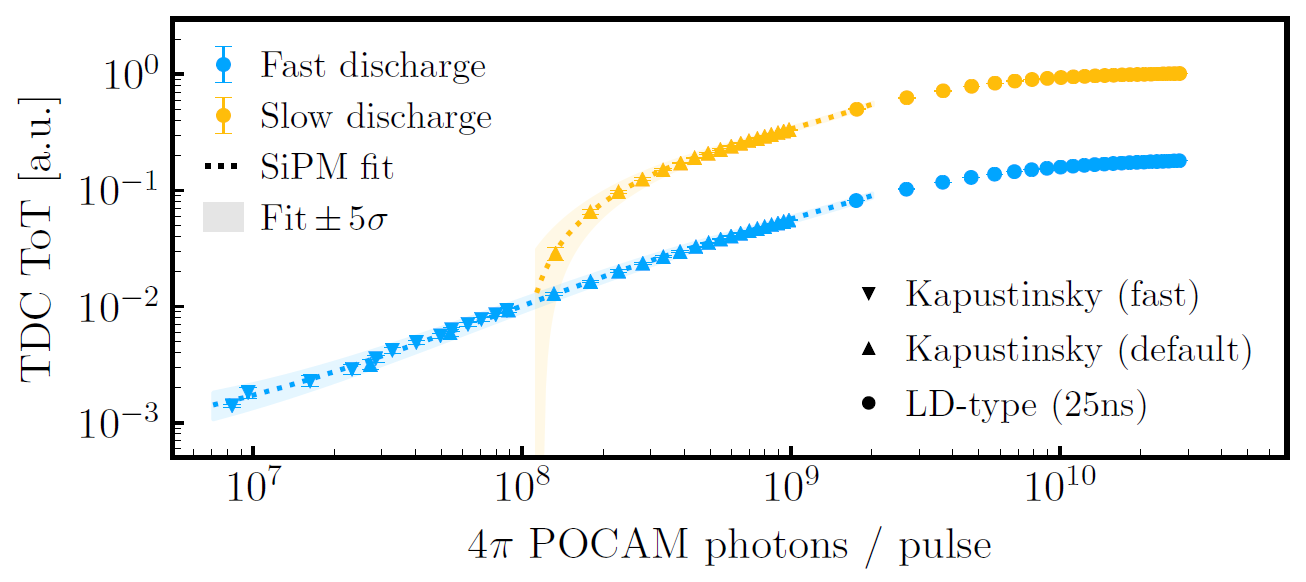}  
%  \label{fig:sub-second}
\end{subfigure}
\caption{\textit{(Left) Readout response of the photodiode for all the drivers at 405 nm.(Right) Readout response of the SiPM for all the drivers at 405 nm.(SiPM over-voltage 5V). \cite{3abc}}}
\label{fig6}
\end{figure}

%\begin{figure}[H]
%	\centering
%	\includegraphics[height=5.5cm, width=12.5cm]{sensorchar1.png}
%	\caption{\textit{Readout response of the photodiode for all the drivers at 405 nm.}}
%	\label{fig6}
%	\end{figure}	

The optimized response behaviour of both these sensors as a function of the POCAM light emission is shown in Figure \ref{fig6}. As expected, the PD showed linear behaviour across all its measurable range. The curve shows saturation at the high gain channel but as can be seen is compensated by the low gain channel. Also the expected saturation curve of the SiPM can easily be fit with appropriate exponential functions \cite{3abc}.

%\begin{figure}[H]
%	\centering
%	\includegraphics[height=5.5cm, width=12.5cm]{sensorchar2.png}
%	\caption{\textit{Readout response of the SiPM for all the drivers at 405 nm.(SiPM over-voltage 5V)}}
%	\label{fig7}
%	\end{figure}	
%	

\section{Calibration setups}
To characterize the flashers and the emission profile of both the hemispheres of the POCAM, two calibration setups have been developed. 
\subsection{Flasher calbration setup}
\begin{figure}[H]
\begin{subfigure}{.55\textwidth}
  \centering
  % include first image
  \includegraphics[width=.8\linewidth]{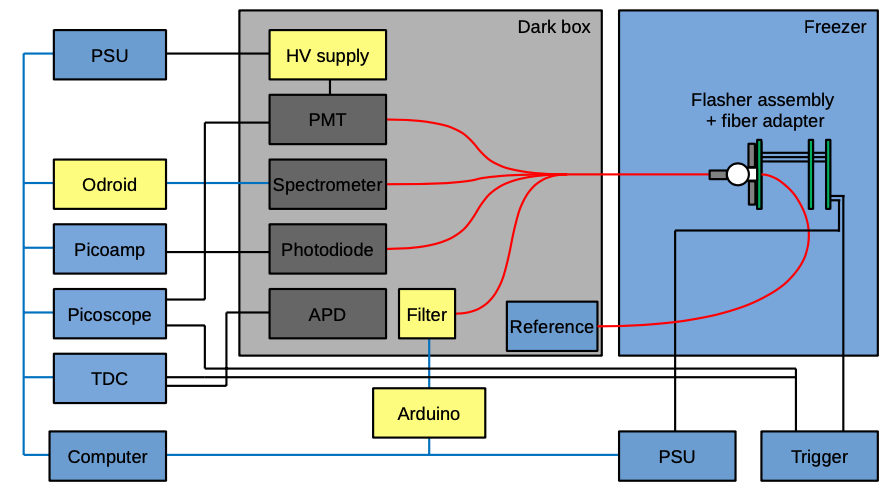}  
  \label{fig:sub-first}
\end{subfigure}
\begin{subfigure}{.45\textwidth}
  \centering
  % include second image
  \includegraphics[width=.8\linewidth]{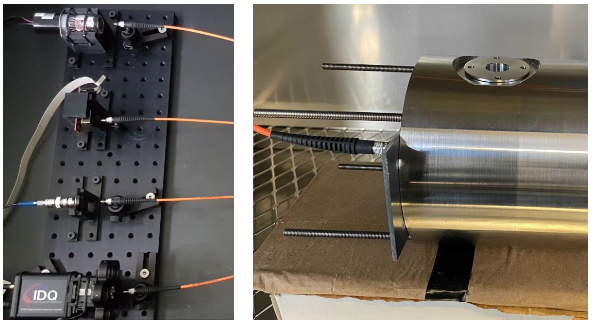}  
  \label{fig:sub-second}
\end{subfigure}
\caption{\textit{Flasher calibration workflow diagram(left) \cite{3abc} and real laboratory setup of the dark box - freezer assembly}}
\label{fig8}
\end{figure}

In the first setup (Figure \ref{fig8}), to calibrate the flashers, a dark box has been built to house sensors which measure characteristics  of the light from the POCAM light source (or any light source), excluding/cutting off the light from the surroundings/natural light. These sensors are each coupled to one end of a 4-to-1 fan-out fiber. This dark box houses four sensors. One, a \textit{photodiode}, which records the light intensity and is read out by a Keithley 6485 pico-ammeter. Next, a \textit{photomultiplier tube (PMT)}, which also records light intensity and the PMT pulses are further recorded with the help of a digital oscilloscope. Third, an \textit{APD}, which uses time-correlated single photon counting to measure the pulse time profile. Moreover, the APD is also equipped with a controllable neutral density filter wheel to achieve low occupancies of below 10 \% which is necessary to provide proper single-photon sampling of the time profile. The APD signal is fed into a high-precision TDC. Finally, a \textit{spectrometer} is also installed, which directly records and outputs the spectrum via serial command.

All of these sensors are connected to the necessary power supplies and other peripheral electronics, and are further  controlled by a dedicated computer running all the necessary software. To begin, the POCAM is first put inside the freezer and the fan-out fiber is coupled to the teflon sphere. This system then performs automatic temperature scans and measures the pulse properties along with the self-monitoring sensor responses. The data generated is then used offline for characterization of individual POCAM hemispheres (upto 3 days per POCAM) giving us fingerprint-characterized or golden POCAMs. 

\begin{figure}[H]
\begin{subfigure}{.60\textwidth}
  \centering
  % include first image
  \includegraphics[width=\textwidth]{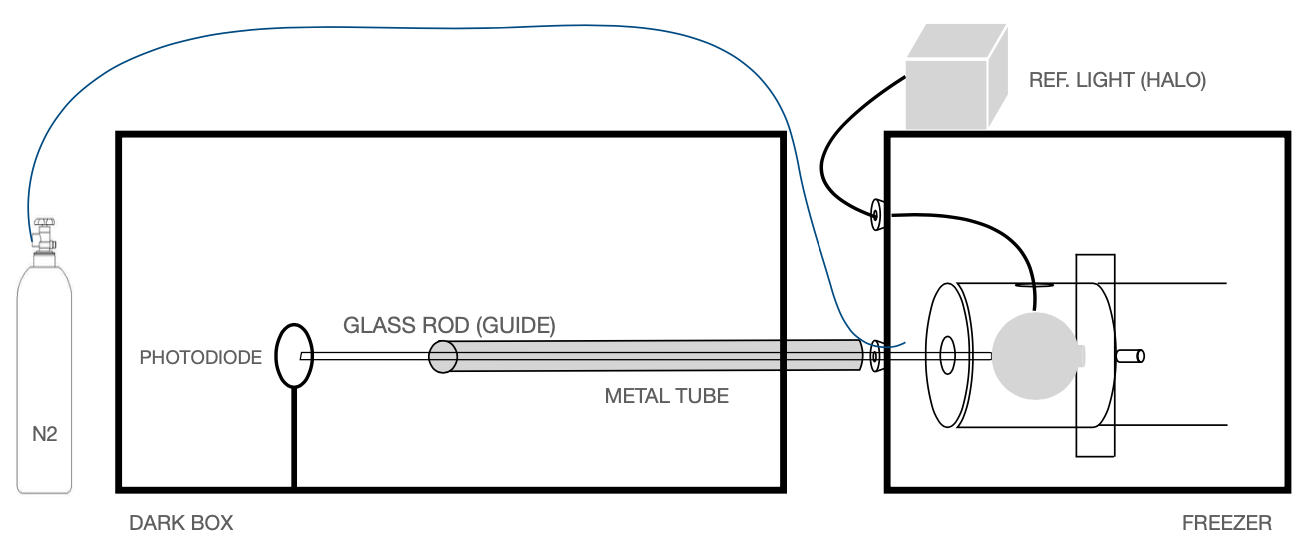}\hspace{1em}
%  \label{fig:sub-first}
\end{subfigure}
\begin{subfigure}{.37\textwidth}
  \centering
  % include second image
  \includegraphics[width=\textwidth]{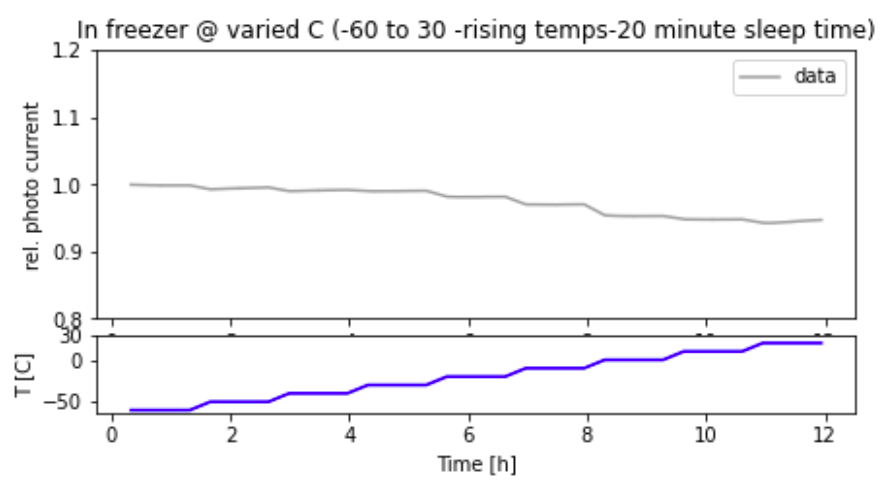}  
%  \label{fig:sub-second}
\end{subfigure}
\caption{\textit{(Left) Schematic of lab setup for reference source monitoring.(Right) Reference source intensity as a function of time. The sleep time here is the duration the freezer takes to gets stable between temperature changes. A quartz glass guide was used to guide the light from the integrating sphere to the sensor}}
\label{fig9}
\end{figure}

\textit{Nitrogen flushing}: To remove further systematic uncertainties resulting from potential fiber coupling changes over the course of cooling and heating, we flush the freezer with nitrogen. Moreover we also monitor a temperature-stabilized reference halogen light source coupled into the integrating sphere using the same type of fiber. A schematic of our current setup is shown in Figure \ref{fig9}. With this setup, we monitored only our reference source, to see any temperature dependent intensity variations. For this purpose, an additional 3-D printed holder for the photodiode is also placed inside the dark box. For this particular procedure, the photodiode needs to be mounted on this holder rather than sit back with the other sensors. It is possible to adjust the position of the holder with the help of screws on the floor of the dark box. The height however is unchangeable because it  is set so as to see the light from the light source in the freezer. Carrying out this measurement, the plot in Figure \ref{fig9} was generated.

\subsection{Relative isotropy calibration}
\begin{figure}[H]
	\centering
	\includegraphics[height=5.5cm, width=12.5cm]{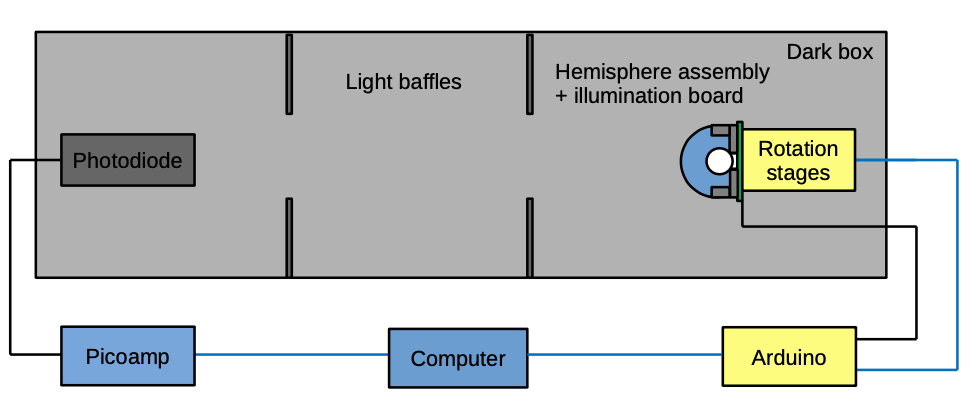}
	\caption{\textit{Schematic of the emission profile measurement station \cite{3abc}}}
	\label{fig10}
	\end{figure}

The second calibration station is made to measure the emission profile of the POCAM (Figure \ref{fig10}). This setup also consists of a dark box, which has a two-axis rotation stage assembly, where the POCAM hemisphere (i.e the flange, integrating spheres, apperture disk and sensors) can be mounted. At the other end of the dark box, just opposite the rotation stage , a photodiode is mounted. The light baffles in between further reduce stray light from reflections off of inner surfaces. After mounting the POCAM hemisphere on the rotation stage, light from the LEDs on the simplified analog board (with a layout identical to the analog board i.e the same led matrix and light emission characteristics) then falls on the PD. A dedicated computer, running all the necessary software, then carries out the azimuth and zenith angles scans, runs the LEDs and measures the intensity seen by the PD. All of this data which is eventually written to file, can then be accessed offline. The rotation of the hemisphere provides a relative characterization of its emission profile and can further be used to calculate the total hemispherical light yield. 

\textit{Absolute calibration}: After successfully finishing the relative calibration of all POCAM hemispheres, an absolute calibration using a NIST (National Institute for Standards and Technology) calibrated PD is carried out. These absolute intensity scales then act as a reference for self-monitoring data over the course of the operational period of the POCAM and can be used for correcting the instrument emission in-situ using the integrated photosensors. An absolute light yield uncertainty of $4.1 \%$ is anticipated, the dominant uncertainty coming from temperature-dependent fiber coupling ($2 \%$).

\section{POCAM prototype calibration}
\begin{figure}
\begin{subfigure}{.49\textwidth}
  \centering
  % include first image
  \includegraphics[height=3.5cm, width = 7cm]{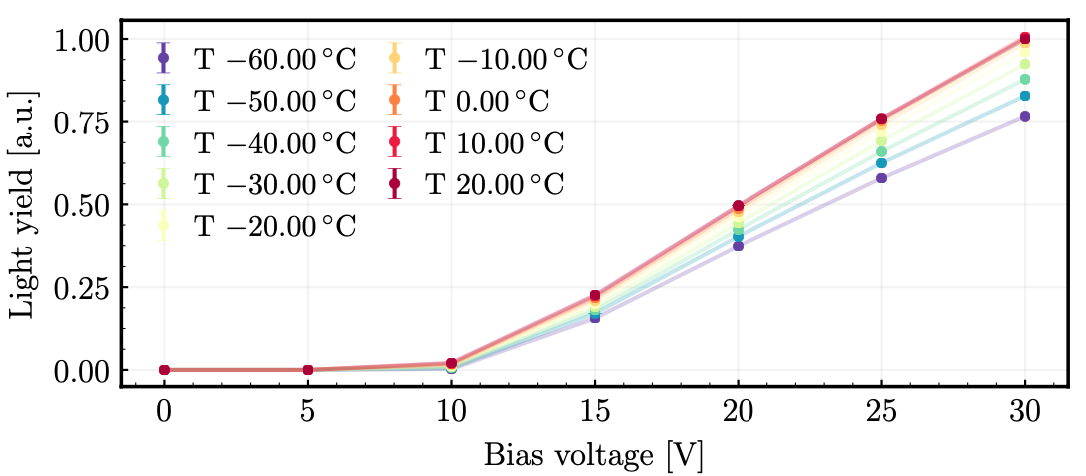}\hspace{1em}
%  \label{fig:sub-first}
\end{subfigure}
\begin{subfigure}{.49\textwidth}
  \centering
  % include second image
  \includegraphics[height=3.5cm, width = 7cm]{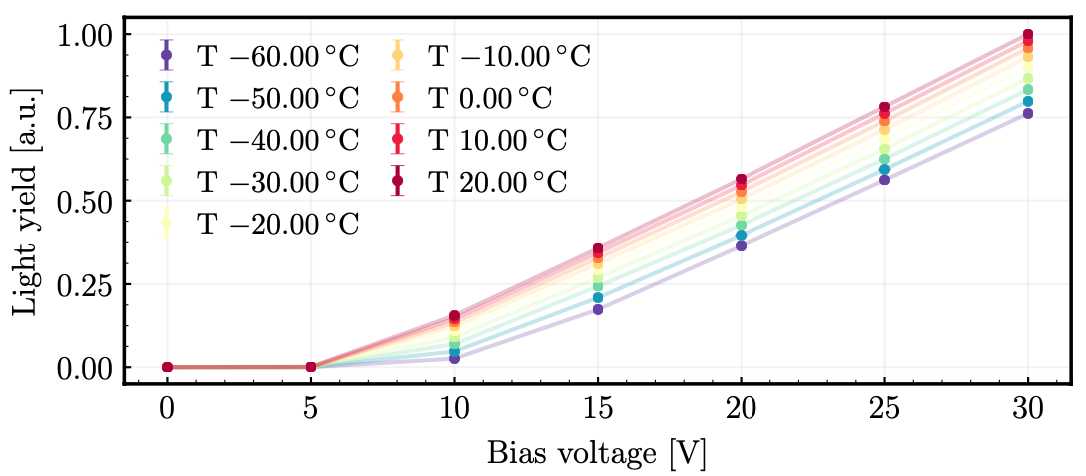}  
%  \label{fig:sub-second}
\end{subfigure}
\caption{\textit{Pulser intensity as a function of voltage at different temperatures for the default Kapustinsky (left) and the LD driver at 25ns (right) \cite{3abc}}}
\label{fig11}
\end{figure}

Both flasher and emission profile characterizations were performed on a prototype POCAM before going towards production. The pulser intensity was measured for all driver types, using the 405nm emitter (Figure \ref{fig11}). The pulsers show lower intensities at lower temperatures, which could possibly be reasoned by the increased value of the series resistance.The time profile and emission spectrum of the light were also measured but they did not show a significant temperature dependence at high intensities. At lower intensities, however, the time profile plot becomes longer for the LD driver, but it shouldn't be a problem since the LD drivers will be used predominantly for high intensities.

\begin{figure}[H]
\begin{subfigure}{.49\textwidth}
  \centering
  % include first image
  \includegraphics[height=3cm, width = 7cm]{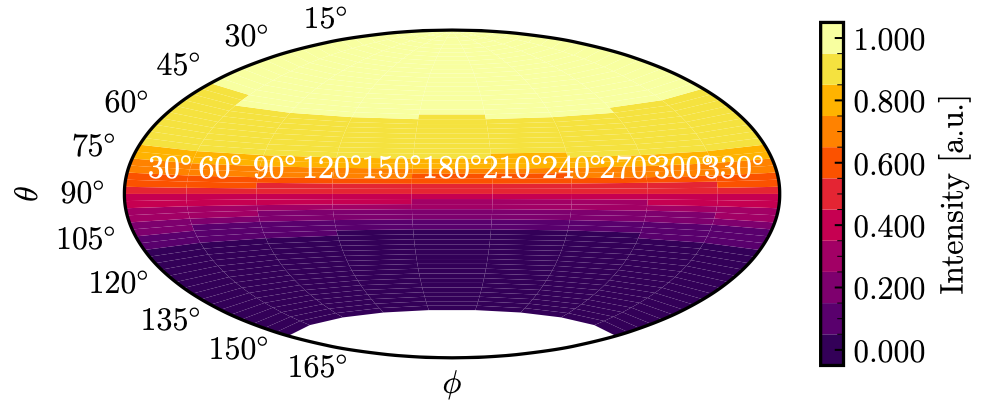}\hspace{1em}
%  \label{fig:sub-first}
\end{subfigure}
\begin{subfigure}{.49\textwidth}
  \centering
  % include second image
  \includegraphics[height=3cm, width = 7cm]{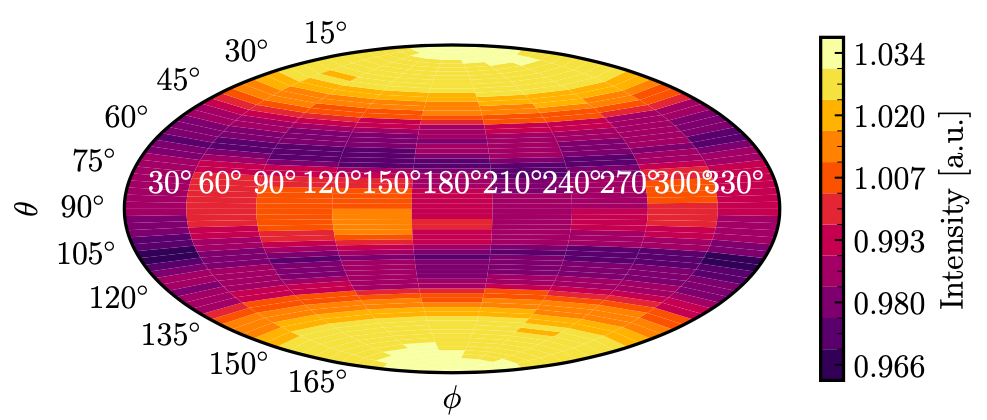}  
%  \label{fig:sub-second}
\end{subfigure}
\caption{\textit{Prototype POCAM single hemisphere emission profile (left) and virtual emission profile of the complete POCAM estimated by overlapping two single hemisphere emissions(right) \cite{3abc}}}
\label{fig12}
\end{figure}

For isotropy measurements, the emission profile of the prototype POCAM hemisphere and then the virtual sum of two hemispheres so as to replicate that of a complete POCAM, is shown via a Mollweide projection in Figure \ref{fig12}. 
By averaging per LED and all azimuthal angles and using the resulting standard deviation as errors, the isotropy measured shows a
$1\sigma$ -error of $1.5 \%$ over the entire zenith range and only $0.4\%$ between $ 0 - 60 \deg$.
Estimation of the total $4\pi $ light yield with 405nm LED gives a typical value of $(5.1 \pm 0.4) \times 10^7$ photons per pulse for the fast Kapustinsky, $(7.5 \pm 0.6) \times 10^8$ photons per pulse for the default Kapustinsky and $(2.4 \pm 0.2) \times 10^{10}$ photons per pulse for the LD type at 25 ns. \cite{3abc}

% Full authors list (ONLY FOR COLLABORATIONS)
\clearpage
\section*{Full Author List: IceCube Collaboration}

% \noindent \textbf{Note comment afterwards:} Collaborations have the possibility to provide an authors list in xml format which will be used while generating the DOI entries making the full authors list searchable in databases like Inspire HEP. For instructions please go to icrc2021.desy.de/proceedings or contact us under icrc2021proc@desy.de.\\

% \scriptsize
% \noindent
% first.author$^1$, 
% second.author$^2$, 
% third.author$^3$ % .... more names
% and 
% last.author$^{n}$ \\

% \noindent
% $^1$first.affiliation.
% $^2$second.affiliation. % .... more affiliation
% $^{m}$last.affiliation.

\scriptsize
\noindent
R. Abbasi$^{17}$,
M. Ackermann$^{59}$,
J. Adams$^{18}$,
J. A. Aguilar$^{12}$,
M. Ahlers$^{22}$,
M. Ahrens$^{50}$,
C. Alispach$^{28}$,
A. A. Alves Jr.$^{31}$,
N. M. Amin$^{42}$,
R. An$^{14}$
% ...
\\

\noindent
% ...
$^{12}$ Universit{\'e} Libre de Bruxelles, Science Faculty CP230, B-1050 Brussels, Belgium \\
% ...
$^{14}$ Department of Physics and Laboratory for Particle Physics and Cosmology, Harvard University, Cambridge, MA 02138, USA \\
% ...
$^{17}$ Department of Physics, Loyola University Chicago, Chicago, IL 60660, USA \\
$^{18}$ Dept. of Physics and Astronomy, University of Canterbury, Private Bag 4800, Christchurch, New Zealand \\
% ...
$^{22}$ Niels Bohr Institute, University of Copenhagen, DK-2100 Copenhagen, Denmark \\
% ...
$^{28}$ D{\'e}partement de physique nucl{\'e}aire et corpusculaire, Universit{\'e} de Gen{\`e}ve, CH-1211 Gen{\`e}ve, Switzerland \\
% ...
$^{31}$ Karlsruhe Institute of Technology, Institute for Astroparticle Physics, D-76021 Karlsruhe, Germany  \\
% ...
$^{42}$ Bartol Research Institute and Dept. of Physics and Astronomy, University of Delaware, Newark, DE 19716, USA \\
% ...
$^{50}$ Oskar Klein Centre and Dept. of Physics, Stockholm University, SE-10691 Stockholm, Sweden \\
% ...
$^{59}$ DESY, D-15738 Zeuthen, Germany \\
% ...

\subsection*{Acknowledgements}

\noindent
USA {\textendash} U.S. National Science Foundation-Office of Polar Programs,
U.S. National Science Foundation-Physics Division,
U.S. National Science Foundation-EPSCoR,
Wisconsin Alumni Research Foundation,
Center for High Throughput Computing (CHTC) at the University of Wisconsin{\textendash}Madison,
Open Science Grid (OSG),
Extreme Science and Engineering Discovery Environment (XSEDE),
Frontera computing project at the Texas Advanced Computing Center,
U.S. Department of Energy-National Energy Research Scientific Computing Center,
Particle astrophysics research computing center at the University of Maryland,
Institute for Cyber-Enabled Research at Michigan State University,
and Astroparticle physics computational facility at Marquette University;
Belgium {\textendash} Funds for Scientific Research (FRS-FNRS and FWO),
FWO Odysseus and Big Science programmes,
and Belgian Federal Science Policy Office (Belspo);
Germany {\textendash} Bundesministerium f{\"u}r Bildung und Forschung (BMBF),
Deutsche Forschungsgemeinschaft (DFG),
Helmholtz Alliance for Astroparticle Physics (HAP),
Initiative and Networking Fund of the Helmholtz Association,
Deutsches Elektronen Synchrotron (DESY),
and High Performance Computing cluster of the RWTH Aachen;
Sweden {\textendash} Swedish Research Council,
Swedish Polar Research Secretariat,
Swedish National Infrastructure for Computing (SNIC),
and Knut and Alice Wallenberg Foundation;
Australia {\textendash} Australian Research Council;
Canada {\textendash} Natural Sciences and Engineering Research Council of Canada,
Calcul Qu{\'e}bec, Compute Ontario, Canada Foundation for Innovation, WestGrid, and Compute Canada;
Denmark {\textendash} Villum Fonden and Carlsberg Foundation;
New Zealand {\textendash} Marsden Fund;
Japan {\textendash} Japan Society for Promotion of Science (JSPS)
and Institute for Global Prominent Research (IGPR) of Chiba University;
Korea {\textendash} National Research Foundation of Korea (NRF);
Switzerland {\textendash} Swiss National Science Foundation (SNSF);
United Kingdom {\textendash} Department of Physics, University of Oxford.
\end{document}